\documentclass[aps,prd,onecolumn,groupedaddress,showpacs,nofootinbib,amssymb]{revtex4}
\usepackage[dvips]{graphicx}
\usepackage{amssymb}
\usepackage{amsmath}
\usepackage{graphicx,color}
\usepackage{amsfonts}
\usepackage{bm}
\usepackage{cancel}
\usepackage{comment}
\usepackage{hyperref}

\newcommand{\e}{\mathrm{e}}

\allowdisplaybreaks[4]

\begin{document}

\tolerance=5000

\title{Propagation Speed of Gravitational Wave in Scalar--Einstein--Gauss-Bonnet Gravity}

\author{Shin'ichi~Nojiri$^{1,2,3}$\thanks{nojiri@gravity.phys.nagoya-u.ac.jp}, 
Sergei.~D.~Odintsov$^{3,4}$\thanks{odintsov@ieec.cat}}

\affiliation{ $^{1)}$ Department of Physics, Nagoya University,
Nagoya 464-8602, Japan \\
$^{2)}$ Kobayashi-Maskawa Institute for the Origin of Particles
and the Universe, Nagoya University, Nagoya 464-8602, Japan \\
$^{3)}$ Institute of Space Sciences (ICE, CSIC) C. Can Magrans
s/n, 08193 Barcelona, Spain \\
$^{4)}$ ICREA, Passeig Luis Companys, 23, 08010 Barcelona, Spain
}

\begin{abstract}
The propagation speed of the gravitational wave in scalar--Einstein--Gauss-Bonnet (sEGB) gravity is generally different 
from that of light. Using differential equation conditions for the speed of gravitational waves to coincide with the light speed 
in the expanding universe, we constructed a general class of sEGB gravities where this condition is satisfied and realistic inflation occurs.
It is demonstrated that the condition that the speed of gravitational wave coincides with that of the light in the Friedmann-Lema\^{i}tre-Robertson-Walker (FRLW) universe 
is always different from the condition for gravitational wave speed in the sEGB black hole background. 
Moreover, it is shown that when gravitational wave speed in sEGB black hole is equal to the speed of light the black hole spacetime geometry 
is changing too so that formally there is no solution for such sEGB black hole. 
This may indicate that sEGB black holes hardly can be considered as realistic black holes unless some reasonable scenario to make gravitational wave speed 
to be equal to that of light is proposed, at least asymptotically.

\end{abstract}


\maketitle

\section{Introduction}

Scalar--Einstein--Gauss-Bonnet (sEGB) gravity sometimes also called string-inspired modified Gauss-Bonnet gravity is deeply related to string/M-theory. 
It found numerous applications in cosmology and astrophysics (for general review, see \cite{Nojiri:2010wj, Nojiri:2017ncd}).
It is known that it yields a realistic description of the universe's evolution as well as to detectable gravitational wave signal~\cite{Hwang:2005hb, Nojiri:2006je, Nojiri:2005vv, 
Satoh:2007gn, Yi:2018gse, Guo:2009uk, Jiang:2013gza, Kanti:2015pda, 
vandeBruck:2017voa, Kanti:1998jd, Pozdeeva:2020apf, Pozdeeva:2021iwc, Koh:2014bka, Bayarsaikhan:2020jww, Chervon:2019sey, 
DeLaurentis:2015fea, Nozari:2017rta, Odintsov:2018zhw, Kawai:1998ab, Yi:2018dhl, vandeBruck:2016xvt, Kleihaus:2019rbg, 
Bakopoulos:2019tvc, Maeda:2011zn, Bakopoulos:2020dfg, Ai:2020peo, Odintsov:2020xji, Oikonomou:2020sij, Odintsov:2020zkl, 
Odintsov:2020mkz, Venikoudis:2021irr, Kong:2021qiu, Easther:1996yd, Antoniadis:1993jc, Antoniadis:1990uu, Kanti:1995vq, 
Kanti:1997br, Easson:2020mpq, Rashidi:2020wwg, Odintsov:2023aaw, Odintsov:2023lbb}, 
which may even result in a blue-tilted tensor spectrum~\cite{Oikonomou:2022xoq}. 
Such theories have a problem due to a gravitational wave speed different from that of light in the vacuum, (see GW170817 
event works~\cite{TheLIGOScientific:2017qsa, Monitor:2017mdv, GBM:2017lvd}).This looks like some indication of an incorrect description of nature. 
However, the above problem may be somehow solved if the propagation speed of the inflationary gravitational waves is equal to unity in natural units 
(equal to that of light in vacuum and natural units $c=1$).
This can be achieved if their Gauss-Bonnet coupling potential, often denoted as $\xi (\phi)$, is 
constrained via additional equation~\cite{Odintsov:2020xji, Odintsov:2020sqy, Oikonomou:2021kql, Oikonomou:2022ksx}. 

Although the condition that the speed of gravitational wave coincides with that of the light in the Friedmann-Lema\^{i}tre-Robertson-Walker (FLRW) universe 
has been obtained, in this paper, we show that the two speeds are always different from each other in the sEGB black hole background. 
The change in the propagating speed generates lensing effects in addition to standard gravitational lensing. 
Moreover, it looks like the condition to have gravitational wave speed in the sEGB black hole be equal to the speed of light is not compatible 
with the explicit solution of the corresponding sEGB black hole. 
This may indicate that black holes in sEGB gravity are not compatible with GW170817. 
In other words, unless a novel scenario for the realisation of such black holes is found, black holes in sEGB gravity cannot be considered realistic objects.

In the next section, we give the general equation describing the gravitational wave of the sEGB gravity and find the condition that 
the propagation speed of the gravitational wave is identical to that of light in the general background. 
In Section~\ref{FLRW}, by focusing on the FLRW spacetime, we reconsider the condition found in \cite{Odintsov:2020xji, Odintsov:2020sqy, Oikonomou:2021kql, Oikonomou:2022ksx}.
We construct the general sEGB model satisfying the condition. 
For a large class of sEGB gravities, it is demonstrated that realistic inflation occurs. 
In Section~\ref{ssss}, we investigate the propagation speed of the gravitational wave in static and spherically symmetric spacetime, including black holes. 
After we explain how we can construct general static and spherically symmetric spacetime, we find the condition for the propagation speed and show that the condition cannot be satisfied 
in the sEGB black hole spacetime. 
The last section is devoted to the summary and discussion. 

\section{Gravitational Wave in Scalar--Einstein--Gauss-Bonnet gravity}\label{GWGB}

The action of the sEGB gravity in four spacetime dimensions is given by,
\begin{align}
\label{g2}
\mathcal{S}=\int d^4 x \sqrt{-g}\left\{ \frac{1}{2\kappa^2}R
 - \frac{1}{2} \partial_\mu \phi \partial^\mu \phi{ -}V(\phi) - \xi(\phi) \mathcal{G} \right\}\, .
\end{align}
Here $\phi$ is the scalar field and $V(\phi)$ is the potential for $\phi$ and $\xi(\phi)$ is also a function of $\phi$.
Furthermore, $\mathcal{G}$ is the Gauss-Bonnet invariant defined by
\begin{align}
\label{eq:GB}
\mathcal{G} = R^2-4R_{\alpha \beta}R^{\alpha \beta}+R_{\alpha \beta \rho \sigma}R^{\alpha \beta \rho \sigma}\, .
\end{align}
By the variation of the action (\ref{g2}) with respect to the metric $g_{\mu\nu}$, we obtain 
\begin{align}
\label{gb4bD4}
0= &\, \frac{1}{2\kappa^2}\left(- R_{\mu\nu} + \frac{1}{2} g_{\mu\nu} R\right)
+ \frac{1}{2} \partial_\mu \phi \partial_\nu \phi - \frac{1}{4}g_{\mu\nu} \partial_\rho \phi \partial^\rho \phi 
 - \frac{1}{2} g_{\mu\nu}V(\phi) \nonumber \\
&\, - 2 \left( \nabla_\mu \nabla_\nu \xi(\phi)\right)R + 2 g_{\mu\nu} \left( \nabla^2\xi(\phi)\right)R + 4 \left( \nabla_\rho \nabla_\mu \xi(\phi)\right)R_\nu^{\ \rho}
+ 4 \left( \nabla_\rho \nabla_\nu \xi(\phi)\right)R_\mu^{\ \rho} \nonumber \\
&\, - 4 \left( \nabla^2 \xi(\phi) \right)R_{\mu\nu} - 4g_{\mu\nu} \left( \nabla_{\rho} \nabla_\sigma \xi(\phi) \right) R^{\rho\sigma}
+ 4 \left(\nabla^\rho \nabla^\sigma \xi(\phi) \right) R_{\mu\rho\nu\sigma}\, .
\end{align}
Here we have used the following identity which holds true in four dimensions: $0 = \frac{1}{2}g^{\mu\nu} \mathcal{G} -2 R R^{\mu\nu}
 - 4 R^\mu_{\ \rho} R^{\nu\rho} -2 R^{\mu\rho\sigma\tau}R^\nu_{\ \rho\sigma\tau} +4 R^{\mu\rho\nu\sigma}R_{\rho\sigma}$. 
If the spacetime dimensions are larger than four, there appear the terms 
$- \xi(\phi) \left\{ \frac{1}{2}g^{\mu\nu} \mathcal{G} -2 R R^{\mu\nu} - 4 R^\mu_{\ \rho} R^{\nu\rho}
 -2 R^{\mu\rho\sigma\tau}R^\nu_{\ \rho\sigma\tau} +4 R^{\mu\rho\nu\sigma}R_{\rho\sigma} \right\}$ in Eq.~(\ref{gb4bD4}). 
We should note that the Gauss-Bonnet invariant (\ref{eq:GB}) is a total derivative in four dimensions. 
Therefore if $\xi(\phi)$ is a constant, the Gauss-Bonnet term in the action (\ref{g2}) does not give any contribution. 
If there remain the terms with $\xi(\phi)$ which do not include the derivative in Eq.~(\ref{gb4bD4}), the terms contribute 
to the equation even if $\xi(\phi)$ is a constant, which is an explicit confliction. 

For the general variation of the metric
\begin{align}
\label{variation1}
g_{\mu\nu}\to g_{\mu\nu} + h_{\mu\nu}\, ,
\end{align}
we have the following formulae in the leading order of $h_{\mu\nu}$, 
\begin{align}
\label{variation2}
\delta\Gamma^\kappa_{\mu\nu} =&\, \frac{1}{2}g^{\kappa\lambda}\left(
\nabla_\mu h_{\nu\lambda} + \nabla_\nu h_{\mu\lambda} - \nabla_\lambda h_{\mu\nu}
\right)\, ,\nonumber \\
\delta R^\mu_{\ \nu\lambda\sigma}=&\, \nabla_\lambda \delta\Gamma^\mu_{\sigma\nu} 
 - \nabla_\sigma \delta \Gamma^\mu_{\lambda\nu}\, ,\nonumber \\
\delta R_{\mu\nu\lambda\sigma}=&\, \frac{1}{2}\left[\nabla_\lambda \nabla_\nu h_{\sigma\mu} 
 - \nabla_\lambda \nabla_\mu h_{\sigma\nu} 
 - \nabla_\sigma \nabla_\nu h_{\lambda\mu} 
 + \nabla_\sigma \nabla_\mu h_{\lambda\nu} 
+ h_{\mu\rho} R^\rho_{\ \nu\lambda\sigma} 
 - h_{\nu\rho} R^\rho_{\ \mu\lambda\sigma} \right] \, ,\nonumber \\
\delta R_{\mu\nu} =& \frac{1}{2}\left[\nabla^\rho\left(\nabla_\mu h_{\nu\rho} 
+ \nabla_\nu h_{\mu\rho}\right) - \nabla^2 h_{\mu\nu} 
 - \nabla_\mu \nabla_\nu \left(g^{\rho\lambda}h_{\rho\lambda}\right)\right] \nonumber \\
=&\, \frac{1}{2}\left[\nabla_\mu\nabla^\rho h_{\nu\rho} 
+ \nabla_\nu \nabla^\rho h_{\mu\rho} - \nabla^2 h_{\mu\nu} 
 - \nabla_\mu \nabla_\nu \left(g^{\rho\lambda}h_{\rho\lambda}\right) 
 - 2R^{\lambda\ \rho}_{\ \nu\ \mu}h_{\lambda\rho} 
+ R^\rho_{\ \mu}h_{\rho\nu} + R^\rho_{\ \mu}h_{\rho\nu} \right]\, ,\nonumber \\
\delta R =&\, -h_{\mu\nu} R^{\mu\nu} + \nabla^\mu \nabla^\nu h_{\mu\nu} 
 - \nabla^2 \left(g^{\mu\nu}h_{\mu\nu}\right)\, . 
\end{align}
We do not consider the perturbation of the scalar field $\phi$ because we are now interested in the gravitational wave, which is massless and spin-two mode. 
As long as we consider the leading order of the perturbation, the massless spin-two mode does not mix with the scalar mode, which is a massive spin-zero mode 
although the second-order perturbation of the scalar field plays the role of the source of the gravitational wave. 

The variation of (\ref{gb4bD4}) is given by, 
\begin{align}
\label{gb4bD4B}
0=&\, \left\{ \frac{1}{4\kappa^2} R - \frac{1}{2} \partial_\rho \phi \partial_\rho \phi - \frac{1}{2} V(\phi) 
 - 2 \left( \nabla^2\xi(\phi)\right)R - 4 \left( \nabla_{\rho} \nabla_\sigma \xi(\phi) \right) R^{\rho\sigma} \right\} h_{\mu\nu} \nonumber \\
&\, + \bigg\{ \frac{1}{4}g_{\mu\nu} \partial^\tau \phi \partial^\eta \phi - 2 g_{\mu\nu} \left( \nabla^\tau \nabla^\eta \xi(\phi)\right)R 
 - 4 \left( \nabla^\tau \nabla_\mu \xi(\phi)\right)R_\nu^{\ \eta} - 4 \left( \nabla^\tau \nabla_\nu \xi(\phi)\right)R_\mu^{\ \eta} 
+ 4 \left( \nabla^\tau \nabla^\eta \xi(\phi) \right)R_{\mu\nu} \nonumber \\
&\, + 4g_{\mu\nu} \left( \nabla^\tau \nabla_\sigma \xi(\phi) \right) R^{\eta\sigma}
+ 4g_{\mu\nu} \left( \nabla_{\rho} \nabla^\tau \xi(\phi) \right) R^{\rho\eta}
 - 4 \left(\nabla^\tau \nabla^\sigma \xi(\phi) \right) R_{\mu\ \, \nu\sigma}^{\ \, \eta}
 - 4 \left(\nabla^\rho \nabla^\tau \xi(\phi) \right) R_{\mu\rho\nu}^{\ \ \ \ \eta}
\bigg\} h_{\tau\eta} \nonumber \\
&\, + \frac{1}{2}\left\{ 2 \delta_\mu^{\ \eta} \delta_\nu^{\ \zeta} \left( \nabla_\kappa \xi(\phi) \right)R
 - 2 g_{\mu\nu} g^{\eta\zeta} \left( \nabla_\kappa \xi(\phi) \right)R
 - 4 \delta_\rho^{\ \eta} \delta_\mu^{\ \zeta} \left( \nabla_\kappa \xi(\phi) \right)R_\nu^{\ \rho}
 - 4 \delta_\rho^{\ \eta} \delta_\nu^{\ \zeta} \left( \nabla_\kappa \xi(\phi) \right)R_\mu^{\ \rho} \right. \nonumber \\
&\, \left. + 4 g^{\eta\zeta} \left( \nabla_\kappa \xi(\phi) \right) R_{\mu\nu} 
+ 4g_{\mu\nu} \delta_\rho^{\ \eta} \delta_\sigma^{\ \zeta} \left( \nabla_\kappa \xi(\phi) \right) R^{\rho\sigma}
 - 4 g^{\rho\eta} g^{\sigma\zeta} \left( \nabla_\kappa \xi(\phi) \right) R_{\mu\rho\nu\sigma}
\right\} g^{\kappa\lambda}\left( \nabla_\eta h_{\zeta\lambda} + \nabla_\zeta h_{\eta\lambda} - \nabla_\lambda h_{\eta\zeta} \right) \nonumber \\
&\, + \left\{ \frac{1}{4\kappa^2} g_{\mu\nu} - 2 \left( \nabla_\mu \nabla_\nu \xi(\phi)\right) + 2 g_{\mu\nu} \left( \nabla^2\xi(\phi)\right) \right\}
\left\{ -h_{\mu\nu} R^{\mu\nu} + \nabla^\mu \nabla^\nu h_{\mu\nu} - \nabla^2 \left(g^{\mu\nu}h_{\mu\nu}\right) \right\} \nonumber \\
&\, + \frac{1}{2}\left\{ \left( - \frac{1}{2\kappa^2} - 4 \nabla^2 \xi(\phi) \right) \delta^\tau_{\ \mu} \delta^\eta_{\ \nu} 
+ 4 \left( \nabla_\rho \nabla_\mu \xi(\phi)\right) \delta^\eta_{\ \nu} g^{\rho\tau} 
+ 4 \left( \nabla_\rho \nabla_\nu \xi(\phi)\right) \delta^\tau_{\ \mu} g^{\rho\eta} 
 - 4g_{\mu\nu} \nabla^\tau \nabla^\eta \xi(\phi) \right\} \nonumber \\
&\, \qquad \times \left\{\nabla_\tau\nabla^\phi h_{\eta\phi} + \nabla_\eta \nabla^\phi h_{\tau\phi} - \nabla^2 h_{\tau\eta} 
 - \nabla_\tau \nabla_\eta \left(g^{\phi\lambda}h_{\phi\lambda}\right) - 2R^{\lambda\ \phi}{\ \eta\ \tau}h_{\lambda\phi} 
+ R^\phi_{\ \tau}h_{\phi\eta} + R^\phi_{\ \tau}h_{\phi\eta} \right\} \nonumber \\
&\, + 2 \left(\nabla^\rho \nabla^\sigma \xi(\phi) \right) 
\left\{ \nabla_\nu \nabla_\rho h_{\sigma\mu} 
 - \nabla_\nu \nabla_\mu h_{\sigma\rho} 
 - \nabla_\sigma \nabla_\rho h_{\nu\mu} 
 + \nabla_\sigma \nabla_\mu h_{\nu\rho} 
+ h_{\mu\phi} R^\phi_{\ \rho\nu\sigma} 
 - h_{\rho\phi} R^\phi_{\ \mu\nu\sigma} \right\} \, .
\end{align}
We now choose a condition to fix the gauge as follows
\begin{align}
\label{gfc}
0=\nabla^\mu h_{\mu\nu}\, .
\end{align}
Because we are interested in the massless spin-two mode, we also impose the following condition, 
\begin{align}
\label{ce}
0=g^{\mu\nu} h_{\mu\nu} \, .
\end{align}
Then Eq.~(\ref{gb4bD4}) is reduced as follows, 
\begin{align}
\label{gb4bD4B}
0=&\, \left\{ \frac{1}{4\kappa^2} R - \frac{1}{2} \partial_\rho \phi \partial_\rho \phi - \frac{1}{2} V(\phi) 
 - 2 \left( \nabla^2\xi(\phi)\right)R - 4 \left( \nabla_{\rho} \nabla_\sigma \xi(\phi) \right) R^{\rho\sigma} \right\} h_{\mu\nu} \nonumber \\
&\, + \bigg\{ \frac{1}{4}g_{\mu\nu} \partial^\tau \phi \partial^\eta \phi - 2 g_{\mu\nu} \left( \nabla^\tau \nabla^\eta \xi(\phi)\right)R 
 - 4 \left( \nabla^\tau \nabla_\mu \xi(\phi)\right)R_\nu^{\ \eta} - 4 \left( \nabla^\tau \nabla_\nu \xi(\phi)\right)R_\mu^{\ \eta} 
+ 4 \left( \nabla^\tau \nabla^\eta \xi(\phi) \right)R_{\mu\nu} \nonumber \\
&\, + 4g_{\mu\nu} \left( \nabla^\tau \nabla_\sigma \xi(\phi) \right) R^{\eta\sigma}
+ 4g_{\mu\nu} \left( \nabla_{\rho} \nabla^\tau \xi(\phi) \right) R^{\rho\eta}
 - 4 \left(\nabla^\tau \nabla^\sigma \xi(\phi) \right) R_{\mu\ \, \nu\sigma}^{\ \, \eta}
 - 4 \left(\nabla^\rho \nabla^\tau \xi(\phi) \right) R_{\mu\rho\nu}^{\ \ \ \ \eta}
\bigg\} h_{\tau\eta} \nonumber \\
&\, + \frac{1}{2}\left\{ 2 \delta_\mu^{\ \eta} \delta_\nu^{\ \zeta} \left( \nabla_\kappa \xi(\phi) \right)R
 - 4 \delta_\rho^{\ \eta} \delta_\mu^{\ \zeta} \left( \nabla_\kappa \xi(\phi) \right)R_\nu^{\ \rho}
 - 4 \delta_\rho^{\ \eta} \delta_\nu^{\ \zeta} \left( \nabla_\kappa \xi(\phi) \right)R_\mu^{\ \rho} \right. \nonumber \\
&\, \left. + 4g_{\mu\nu} \delta_\rho^{\ \eta} \delta_\sigma^{\ \zeta} \left( \nabla_\kappa \xi(\phi) \right) R^{\rho\sigma}
 - 4 g^{\rho\eta} g^{\sigma\zeta} \left( \nabla_\kappa \xi(\phi) \right) R_{\mu\rho\nu\sigma}
\right\} g^{\kappa\lambda}\left( \nabla_\eta h_{\zeta\lambda} + \nabla_\zeta h_{\eta\lambda} - \nabla_\lambda h_{\eta\zeta} \right) \nonumber \\
&\, - \left\{ \frac{1}{4\kappa^2} g_{\mu\nu} - 2 \left( \nabla_\mu \nabla_\nu \xi(\phi)\right) + 2 g_{\mu\nu} \left( \nabla^2\xi(\phi)\right) \right\}
R^{\mu\nu} h_{\mu\nu} \nonumber \\
&\, + \frac{1}{2}\left\{ \left( - \frac{1}{2\kappa^2} - 4 \nabla^2 \xi(\phi) \right) \delta^\tau_{\ \mu} \delta^\eta_{\ \nu} 
+ 4 \left( \nabla_\rho \nabla_\mu \xi(\phi)\right) \delta^\eta_{\ \nu} g^{\rho\tau} 
+ 4 \left( \nabla_\rho \nabla_\nu \xi(\phi)\right) \delta^\tau_{\ \mu} g^{\rho\eta} 
 - 4g_{\mu\nu} \nabla^\tau \nabla^\eta \xi(\phi) \right\} \nonumber \\
&\, \qquad \times \left\{ - \nabla^2 h_{\tau\eta} - 2R^{\lambda\ \phi}_{\ \eta\ \tau}h_{\lambda\phi} 
+ R^\phi_{\ \tau}h_{\phi\eta} + R^\phi_{\ \tau}h_{\phi\eta} \right\} \nonumber \\
&\, + 2 \left(\nabla^\rho \nabla^\sigma \xi(\phi) \right) \left\{ \nabla_\nu \nabla_\rho h_{\sigma\mu} - \nabla_\nu \nabla_\mu h_{\sigma\rho} 
 - \nabla_\sigma \nabla_\rho h_{\nu\mu} + \nabla_\sigma \nabla_\mu h_{\nu\rho} + h_{\mu\phi} R^\phi_{\ \rho\nu\sigma} 
 - h_{\rho\phi} R^\phi_{\ \mu\nu\sigma} \right\} \, .
\end{align}

The observation of GW170817~\cite{LIGOScientific:2017vwq} gives the constraint on the propagation speed $c_\mathrm{GW}$ of the 
gravitational wave as follows, 
\begin{align} 
\label{GWp9} 
\left| \frac{c_\mathrm{GW}^2}{c^2} - 1 \right| < 6 \times 10^{-15}\, .
\end{align}
Here $c$ is the speed of light. 
The expression (\ref{gb4bD4B}) is rather complex because we are considering the curved spacetime. 
If we consider the local Lorentz frame, there could only be the terms including the second derivative of $h_{\mu\nu}$ as in the equation 
describing the propagation of the gravitational wave in the flat spacetime. 
Therefore in order to investigate if the propagating speed $c_\mathrm{GW}$ of the gravitational wave $h_{\mu\nu}$ could be different from 
that of the light $c$, we only need to check the parts including the second derivatives of $h_{\mu\nu}$, 
which is also equivalent to considering the high-energy gravitational wave whose wavelength is much smaller than the curvature radius, 
\begin{align}
\label{second}
I_{\mu\nu} \equiv&\, I^{(1)}_{\mu\nu} + I^{(2)}_{\mu\nu} \, , \nonumber \\
I^{(1)}_{\mu\nu} \equiv&\, \frac{1}{2}\left\{ \left( - \frac{1}{2\kappa^2} - 4 \nabla^2 \xi(\phi) \right) \delta^\tau_{\ \mu} \delta^\eta_{\ \nu} 
+ 4 \left( \nabla_\rho \nabla_\mu \xi(\phi)\right) \delta^\eta_{\ \nu} g^{\rho\tau} 
+ 4 \left( \nabla_\rho \nabla_\nu \xi(\phi)\right) \delta^\tau_{\ \mu} g^{\rho\eta} 
 - 4g_{\mu\nu} \nabla^\tau \nabla^\eta \xi(\phi) \right\} \nabla^2 h_{\tau\eta} \nonumber \\
I^{(2)}_{\mu\nu} \equiv &\, 2 \left(\nabla^\rho \nabla^\sigma \xi(\phi) \right) 
\left\{ \nabla_\nu \nabla_\rho h_{\sigma\mu} - \nabla_\nu \nabla_\mu h_{\sigma\rho} - \nabla_\sigma \nabla_\rho h_{\nu\mu} 
 + \nabla_\sigma \nabla_\mu h_{\nu\rho} \right\} \, .
\end{align}
We should note that $I^{(1)}_{\mu\nu}$ does not change the speed of the gravitational wave from the speed of light. 
On the other hand, $I^{(2)}_{\mu\nu}$ changes the speed of the gravitational wave from that of the light in general, which may violate the constraint (\ref{GWp9}). 
If $\nabla^\rho \nabla^\sigma \xi(\phi)$ is proportional to the metric $g_{\mu\nu}$, $I^{(2)}_{\mu\nu}$ does not change the speed of the gravitational wave from that of the light 
but it is impossible in the general background. 
There could be a chance, however, because we are interested in massless spin-two mode. 

In the next section, we have obtained the general model satisfying the condition (\ref{sgw1}) that the propagation speed of the gravitational wave in the sEGB gravity 
is identical to that of light in the FLRW spacetime. 

\section{Gravitational wave in Friedmann-Lema\^{i}tre-Robertson-Walker universe}\label{FLRW}

To check the effect by $I^{(2)}_{\mu\nu}$, we assume the Friedmann-Lema\^{i}tre-Robertson-Walker (FLRW) universe with a flat spacial part, whose metric is given by 
\begin{align}
\label{FRW}
ds^2= -dt^2 + a(t)^2\sum_{i=1,2,3} \left(dx^i\right)^2 \, .
\end{align} 
Here $t$ is a cosmological time, and $a(t)$ is the scale factor. 
We often use $H\equiv \frac{\dot a}{a}$ which is the Hubble rate. 
Now
\begin{align}
\label{E2}
& \Gamma^t_{ij}= a^2 H \delta_{ij}\, ,\quad \Gamma^i_{jt}=\Gamma^i_{tj}=H\delta^i_{\ j}\, , \nonumber \\
& R_{itjt}= -\left(\dot H + H^2\right)a^2h_{ij}\, ,\quad 
R_{ijkl}= a^4 H^2 \left(\delta_{ik} \delta_{lj} - \delta_{il} \delta_{kj}\right)\, ,\nonumber \\
& R_{tt}=-3\left(\dot H + H^2\right)\, ,\quad R_{ij}= a^2 \left(\dot H + 3H^2\right) \delta_{ij}\, ,\quad 
R= 6\dot H + 12 H^2\, , \quad \mbox{other components}=0\, .
\end{align}
As we are interested in the massless spin-two mode, we further impose the condition $h_{tt}=0$. 
By using (\ref{gfc}) and (\ref{ce}), $h_{ij}$ satisfies the following conditions, 
\begin{align}
\label{GWS2}
0= \sum_{i=1,2,3} h_{ii} \, , \quad 
0= \sum_{i=1,2,3} \partial_i h_{ij}\, .
\end{align}
Then we find 
\begin{align}
\label{I2}
I^{(2)}_{tt}=&\, I^{(2)}_{it}= I^{(2)}_{ti} = 0 \, , \nonumber \\
I^{(2)}_{ij}=&\, 2 \frac{d^2 \xi}{dt^2} \left\{ - {\partial_t}^2 h_{ij} - \dot H \partial_t h_{ij} - H^2 h_{ij} \right\} 
 - 2 a^{-2} H \frac{d\xi}{dt} \left\{ - \triangle h_{ij} + a^2 H \partial_t h_{ij} + a^2 H^2 h_{ij} \right\}\, .
\end{align}
The constraint (\ref{GWp9}) should be imposed only in the late universe where the gravitational waves have been observed and 
we need not require the constraint (\ref{GWp9}) in the early universe. 
There could be two possibilities so that the constraint (\ref{GWp9}) could be satisfied: 
\begin{enumerate}
\item\label{p1} The first possibility is that the parts including the second derivatives of $h_{\mu\nu}$ in $I^{(2)}_{ij}$ should be also proportional to $\nabla^2$ 
at the late universe, which requires, 
\begin{align}
\label{sgw1}
\frac{d^2 \xi}{dt^2} \sim H \frac{d\xi}{dt}\, ,
\end{align}
as found from Eq.~(\ref{I2}). 
Eq.~(\ref{sgw1}) can be solved as 
\begin{align}
\label{sgw2}
\frac{d \xi}{dt} \sim \xi_0 a(t)\, .
\end{align}
Here $\xi_0$ is a constant and $a(t)$ is the scale factor in (\ref{FRW}). 
\item\label{p2} The second possibility is that $I^{(2)}_{ij}$ itself can be neglected in the late universe. 
This requires that $\xi$ goes to a constant or vanishes in the late universe. 
This tells that the Gauss-Bonnet term in the action (\ref{g2}) becomes a total derivative and does not give any contribution to the expansion of the universe 
although the term may become important in the early universe. 
If this possibility is realised, the theory reduces to the scalar-tensor theory at the late universe. 
\end{enumerate}
To investigate how the above possibilities could be satisfied, we need to investigate the structure of $\xi$. 
{ 
Note that some oscillations on the gravitational wave speed spectrum in the sEGB gravity were discussed in \cite{Cai:2015dta, Cai:2015ipa}. 

}

The equations corresponding to the FLRW equations have the following forms, which are given by Eq.~(\ref{gb4bD4}), 
\begin{align}
\label{SEGB3}
0=&\, - \frac{3}{\kappa^2}H^2 + \frac{1}{2}{\dot\phi}^2 + V(\phi) 
+ 24 H^3 \frac{d \xi(\phi(t))}{dt}\, ,\nonumber \\
0=&\, \frac{1}{\kappa^2}\left(2\dot H + 3 H^2 \right) + \frac{1}{2}{\dot\phi}^2 - V(\phi) 
 - 8H^2 \frac{d^2 \xi(\phi(t))}{dt^2} 
 - 16H \dot H \frac{d\xi(\phi(t))}{dt} - 16 H^3 \frac{d \xi(\phi(t))}{dt} \nonumber \\
0=&\, \ddot \phi + 3H\dot \phi + V'(\phi) + \xi'(\phi) \mathcal{G}\, . 
\end{align}
We now use the $e$-folding number $N$ defined by $a=a_0\e^N$ instead of the cosmological time $t$. 
The equations (\ref{SEGB3}) can be integrated by using the $e$-folding number $N$ as follows, 
\begin{align}
\label{SEGB11}
V(\phi(N)) =&\, \frac{3}{\kappa^2}H(N)^2 - \frac{1}{2}H(N)^2 \phi' (N)^2 - 3\e^N H(N) 
\int^N \frac{dN_1}{\e^{N_1}} \left(\frac{2}{\kappa^2} H' (N_1) + H(N_1) \phi'(N_1)^2 \right)\, , \\
\label{SEGB10}
\xi(\phi(N))=&\, \frac{1}{8}\int^N dN_1 \frac{\e^{N_1}}{H(N_1)^3} \int^{N_1} \frac{dN_2}{\e^{N_2}}
\left(\frac{2}{\kappa^2}H' (N_2) + H(N_2) {\phi'(N_2)}^2 \right)\, . 
\end{align}
Eq.~(\ref{SEGB10}) tells that by using functions $h(N)$ and $\tilde f(\phi)$, if $\xi(\phi)$ and $V(\phi)$ are given by 
\begin{align}
\label{SEGB12}
V(\phi) =&\, \frac{3}{\kappa^2}h \left(\tilde f(\phi)\right)^2 
 - \frac{h\left(\tilde f\left(\phi\right)\right)^2}{2\tilde f'(\phi)^2} - 3h\left(\tilde f(\phi)\right) \e^{\tilde f(\phi)} 
\int^\phi d\phi_1 \tilde f'( \phi_1 ) \e^{-\tilde f(\phi_1)} \left(\frac{2}{\kappa^2}h'\left(\tilde f(\phi_1)\right) 
+ \frac{h'\left(\tilde f\left(\phi_1\right)\right)}{f'(\phi_1 )^2} \right)\, , \\
\label{SEGB13}
\xi(\phi) =&\, \frac{1}{8}\int^\phi d\phi_1 
\frac{\tilde f'(\phi_1) \e^{\tilde f(\phi_1)} }{h\left(\tilde f\left(\phi_1\right)\right)^3} 
\int^{\phi_1} d\phi_2 f'(\phi_2) \e^{-\tilde f(\phi_2)} \left(\frac{2}{\kappa^2}h'\left(\tilde f(\phi_2)\right) 
+ \frac{h\left(\tilde f(\phi_2)\right)}{\tilde f'(\phi_2)^2} \right)\, , 
\end{align}
a solution of the equations in (\ref{SEGB3}) is given by 
\begin{align}
\label{SEGB14}
\phi=\tilde f^{-1}(N)\quad \left(N=\tilde f(\phi)\right)\, ,\quad H = h(N) \, .
\end{align}
Therefore we obtain a general sEGB model 
realising the time-evolution of $H$ given by an arbitrary function $h(N)$ as in (\ref{SEGB14}). 

For example, we may consider the following expansion of the universe, 
\begin{align}
\label{phantom1}
H= \frac{h_0}{t_s - t} = \frac{h_0}{t_1} \e^\frac{N}{h_0} \, , \quad 
\phi = \phi_0 \ln \frac{t_s - t}{t_1} = - \frac{\phi_0}{h_0} N \, , \quad 
\left( N=- h_0 \ln \frac{t_s - t}{t_1} \, \quad \tilde f(\phi) = - \frac{h_0 \phi}{\phi_0} \right)\, . 
\end{align}
Here $h_0$, $t_s$, $\phi_0$. and $t_1$ are positive constants. 
Eq.~(\ref{phantom1}) describes the phantom universe, that is, the universe generated by Einstein's gravity coupled with a perfect fluid whose EoS 
parameter $w=\frac{p}{\rho}<-1$. 
Here $p$ and $\rho$ are the pressure and the energy density of the perfect fluid. 
By using (\ref{SEGB12}) and (\ref{SEGB13}), we find that Eq.~(\ref{phantom1}) gives, 
\begin{align}
\label{phantom2}
V(\phi)=&\, V_0 \e^{- \frac{2\phi}{\phi_0}}\, , \quad \xi(\phi) = \xi_0 \e^\frac{2\phi}{\phi_0}\, , \nonumber \\
V_0 =&\, - \frac{1}{\kappa^2 \left( 1 - h_0 \right){t_1}^2}\left\{ 3 {h_0}^2 \left( 1 + h_0 \right) + \frac{{\phi_0}^2 \kappa^2 \left( 1 + 5 h_0 \right)}{2} \right\} \, , \nonumber \\
\xi_0 =&\, \frac{{t_1}^2}{8{h_0}^2 \kappa^2 \left( 1 - h_0 \right)} \left( h_0 + \frac{{\phi_0}^2 \kappa^2}{2} \right) \, .
\end{align}
We should note the phantom universe cannot be realized by the standard canonical scalar-tensor theory with $\xi(\phi)=0$ and with non-negative potential $V(\phi)\geq 0$. 

We should note that the history of the expansion of the universe is determined only by the function $h(N)$ and does not depend on the choice 
of $f(\phi)$. 
By using this indefiniteness of the choice of $f(\phi)$, we consider the two possibilities in \ref{p1} and \ref{p2}. 

For the first possibility in \ref{p1}, by combining (\ref{sgw2}) and (\ref{SEGB10}), we find 
\begin{align}
\label{p1A1}
\xi_0 \e^N = \frac{\e^N}{8H(N)^2} \int^N \frac{dN_1}{\e^{N_1}}
\left(\frac{2}{\kappa^2}H' (N_1) + H(N_1) {\phi'(N_1)}^2 \right)\, . 
\end{align}
Here we used the definition of the $e$-foldings $a=\e^N$ and $H=\frac{dN}{dt}$. 
Eq.~(\ref{p1A1}) can be rewritten as 
\begin{align}
\label{p1A2}
16 \xi_0 H(N) H'(N) = \e^{-N} \left(\frac{2}{\kappa^2}H' (N) + H(N) {\phi'(N)}^2 \right)\, , 
\end{align}
which gives 
\begin{align}
\label{p1A2}
\phi (N) = f^{-1}(N) = \int dN \sqrt{ 16\xi_0 H'(N) \e^N - \frac{2H'(N)}{\kappa^2 H(N)}}\, . 
\end{align}
Therefore the first possibility in \ref{p1} can be also realised by the choice of the function $f(\phi)$. 

Eq.~(\ref{SEGB10}) shows that the second possibility \ref{p2} can be satisfied if 
\begin{align}
\label{p2A1}
0 \sim \frac{2}{\kappa^2}H' (N) + H(N) {\phi'(N)}^2 \, ,
\end{align}
which can be solved as 
\begin{align}
\label{p2A2}
\phi (N) = f^{-1}(N) = \int dN \sqrt{ - \frac{2H'(N)}{\kappa^2 H(N)}}\, . 
\end{align}
The above expression can be valid as long as $H'(N)<0$, which corresponds to the case that the effective equation of state parameter is greater than $-1$. 

If we require that the propagation speed of the gravitational wave always coincides with that of light, we obtain Eqs.~(\ref{sgw2}) and (\ref{p1A2}). 
Eq.~(\ref{p1A2}) gives, 
\begin{align}
\label{dphidN}
\phi'(N)^2 = 16\xi_0 H'(N) \e^N - \frac{2H'(N)}{\kappa^2 H(N)}\, .
\end{align}
Then by using Eqs.~(\ref{SEGB11}) and (\ref{SEGB10}), we obtain 
\begin{align}
\label{SEGB11BB}
V(\phi(N)) =&\, \frac{1}{\kappa^2} \left( 3 H(N)^2 + H(N) H'(N) \right) - 24 \xi_0 \e^N \left( H(N)^3 + H(N)^2 H'(N) \right) \, , \\
\xi'(\phi(N)) =&\, \frac{\xi_0 \e^N}{H(N)\sqrt{ 16\xi_0 H'(N) \e^N - \frac{2H'(N)}{\kappa^2 H(N)} }}\, . 
\end{align}
Thus we have obtained the general sEGB gravity satisfying the condition (\ref{sgw1}) that the propagation speed of the gravitational wave is identical 
to that of light in the FLRW spacetime. 

In the case of the sEGB gravity model, the slow-roll parameters are given by \cite{Nojiri:2023mvi}, 
\begin{align}
\label{slGB1}
\varepsilon_\mathrm{GB} = \frac{\kappa^2}{36} \left( \frac{V_\mathrm{eff}'\left(\phi\right)}{H^2} \right)^2\, , \quad
\eta_\mathrm{GB} = \frac{V_\mathrm{eff}''\left(\phi\right)}{6H^2} \, .
\end{align}
Here
\begin{align}
\label{slGB2}
V_\mathrm{eff}\left(\phi\right) = V(\phi) + \xi(\phi) \mathcal{G} = V(\phi) + 24 \left( H^3 H' + H^4 \right) \xi(\phi) \, .
\end{align}
We now evaluate the above slow-roll parameters. 

We need to speculate about the order of $\zeta_0 \e^N$ in (\ref{sgw2}). 
If we assume $16\xi_0 \e^N \sim \frac{2}{\kappa^2 H(N)} $by using (\ref{dphidN}), at the present universe, $N\sim 120 \sim 140$, 
we find $16 \left| \xi_0 \right| \e^N \ll \frac{2}{\kappa^2 H(N)}$ in the epoch of the inflation $N< 60\sim 70$ and we obtain, 
\begin{align}
\label{SEGB11BB6}
V' (\phi(N)) =&\, \frac{6 H(N) H'(N) + H'(N)^2 + H(N) H''(N)}{\kappa \sqrt{ - \frac{2H'(N)}{H(N)} }} 
\sim \frac{6}{\kappa} \sqrt{- \frac{H(N)^3 H'(N)}{2}} \, , \\
\label{SEGB11BB7}
V'' (\phi(N)) =&\, - \frac{1}{\frac{2 H'(N)}{H(N)}} \left\{- \frac{ \left( 6 H(N) H'(N) + H'(N)^2 + H(N) H''(N) \right) 
\left( H(N) H''(N) - H'(N)^2 \right)}{2 H(N) H'(N) } \right. \nonumber \\
&\, + \left( 6 H'(N)^2 + 6 H(N) H''(N) + 2 H'(N) H''(N) + H'(N) H''(N) + H(N) H'''(N) \right) \biggr\} \nonumber \\
\sim &\, - \frac{3 H(N)^2 H''(N)}{2 H'(N)} \, , \\
\label{SEGB10BB8}
\xi''(\phi(N)) =&\, - \frac{\xi_0 \e^N}{\frac{2H'(N)}{\kappa^2 H(N)}} \left\{ \frac{ H(N) H''(N) - H'(N)^2 }{2 H(N)^2 H'(N)} 
 - \frac{H'(N)}{H(N)^2} \right\} 
\sim - \frac{\kappa^2 \xi_0 \e^N H''(N)}{4 H'(N)^2} \, .
\end{align}
Here we have further assumed $H\gg H' \gg H'' \gg H'''$ and $HH'' \gg {H'}^2$. 

By assuming $H\sim H_0 \left( 1 - \alpha N^\beta \right)$ $\left( \alpha,\, \beta>0\, , \ \alpha N^\beta \ll 1 \right)$, 
the slow-roll parameters are given by, 
\begin{align}
\label{slGB1}
\varepsilon_\mathrm{GB} \sim - \frac{H(N)'}{2H(N)} \sim \frac{\alpha\beta N^{\beta-1}}{2} \, , \quad 
\eta_\mathrm{GB} \sim - \frac{H''(N)}{4 H'(N)} = - \frac{\beta - 1}{4N} \, ,
\end{align}
and therefore we obtain
\begin{align}
\label{slGB2}
r= 16 \varepsilon_\mathrm{GB} \sim 8 \alpha\beta N^{\beta-1}\, , \quad 
n_s = 1 - \frac{\beta - 1}{2N} - 3 \alpha\beta N^{\beta-1} \, .
\end{align}
We should note $\beta$ and $\gamma$ are positive but $\alpha$ is negative. 
We may adjust the parameters to satisfy the Planck 2015 data \cite{Planck:2015sxf}, $r<0.11$ and $n_s = 0.968\pm 0.006$ 
and updated data in the Planck 2018 observation \cite{Planck:2018vyg}, $r<0.06$ and $n_s=0.965\pm 0.004$.

\section{Gravitational wave in Static and Spherically Symmetric Spacetime}\label{ssss}

In general, the constraint (\ref{sgw1}) obtained from the FLRW spacetime is not sufficient to give the condition that the speed of the gravitational wave coincides 
with that of light in the black hole background. 
In the FLRW spacetime, the constraint is obtained by requiring that the ratio of the coefficient $\partial_t^2$ term 
and the coefficient of $\triangle \equiv \partial_i \partial^i$ in the equation for the propagation of the gravitational wave in (\ref{gb4bD4B}) 
should be identical to that of the standard d'Alembertian $\box \equiv \nabla_\mu \nabla^\mu$ given by metric. 
In the case of the black hole background, we need to require that the ratios for the three quantities, that is, the coefficient of $\partial_t^2$, that of $\partial_r^2$, 
and that of the angular components must coincide with those of the d'Alembertian given by metric of the black hole background. 
Therefore the black hole requirements are much stronger than those in the FLRW spacetime and do not always coincide with or include the constraint for the FLRW background.

\subsection{Reconstruction of static and spherically symmetric spacetime}

Before considering the propagation of the gravitational wave in the background of the black hole spacetime, we explain how general static and spherically symmetric spacetime, 
which includes the black hole spacetime, can be constructed in the framework of the sEGB gravity \cite{Nashed:2021cfs, Nojiri:2023qgd}. 

We now assume the spacetime metric has the following form:
\begin{align}
\label{GBiv}
ds^2 = - \e^{2\nu (r)} dt^2 + \e^{2\lambda (r)} dr^2 
+ r^2 \sum_{i,j=1}^2 \tilde g_{ij} dx^i dx^j\ .
\end{align}
Here $\tilde g_{ij}$ is the metric of the two-dimensional sphere with a unit radius. 
In this section, we choose $i,j=1,2$, which is different from the convention $i,j=1,2,3$ in the last section. 

For the metric (\ref{GBiv}), the non-vanishing connections and curvatures are, 
\begin{align}
\label{GBv0}
\Gamma^r_{tt}=&\, \e^{-2\lambda + 2\nu}\nu' \, ,\quad \Gamma^t_{tr}=\Gamma^t_{rt}=\nu'\, ,\quad 
\Gamma^r_{rr}= \lambda'\, ,\quad \Gamma^i_{jk}=\tilde \Gamma^i_{jk}\, ,\quad 
\Gamma^r_{ij}= -\e^{-2\lambda}r\tilde g_{ij}\, ,\nonumber \\
\Gamma^i_{rj}=&\, \Gamma^i_{jr}=\frac{1}{r}\delta^i_{\ j}\, , \quad \Gamma^i_{jk} = {\tilde\Gamma}^i_{jk}\, , \\
\label{GBv}
R_{rtrt}=&\, \e^{2\nu}\left\{\nu'' + \left(\nu' - \lambda'\right) \nu'\right\} \, ,\quad 
R_{titj}= r\nu'\e^{2(\nu - \lambda)} \tilde g_{ij} \, ,\quad 
R_{rirj}= r\lambda' \tilde g_{ij} \, ,\nonumber \\
R_{ijkl}=&\, \left(1 - \e^{-2\lambda}\right) r^2 \left(\tilde g_{ik} \tilde g_{jl} - \tilde g_{il} \tilde g_{jk} \right)\, ,\nonumber \\
R_{tt}=&\, \e^{2\left(\nu - \lambda\right)} \left\{ \nu'' + \left(\nu' - \lambda'\right)\nu' + \frac{2 \nu'}{r}\right\} \, ,\quad 
R_{rr}= - \left\{ \nu'' + \left(\nu' - \lambda'\right)\nu' \right\} + \frac{2 \lambda'}{r} \, ,\nonumber \\
R_{ij}=&\, \left[ k - \left\{ 1 + r \left(\nu' - \lambda' \right)\right\}\e^{-2\lambda}\right] \tilde g_{ij}\, , \nonumber \\
R=&\, \e^{-2\lambda}\left[ - 2\nu'' - 2\left(\nu' - \lambda'\right)\nu' - \frac{4 \left(\nu' - \lambda'\right)}{r} 
+ \frac{2 \left( \e^{2\lambda} - 1 \right)}{r^2} \right] \ .
\end{align}
In (\ref{GBv0}), ${\tilde\Gamma}^i_{jk}$ is the connections of the two-dimensional sphere with a unit radius and 
in (\ref{GBv0}) and (\ref{GBv}), we denote the derivative with respect to $r$ by $'$. 

For the metric given by Eq. (\ref{GBiv}), the $(t,t)$-, $(r,r)$-, and angular components of Eq.~(\ref{gb4bD4}) and the equation for the scalar field $\phi$ 
have the following forms,
\begin{align}
\label{Eq2tt}
0 =&\,- 16 \left(1 - \e^{-2\lambda}\right) \xi'' - 4 \left\{ - 4\left( 1-3\e^{-2\lambda} \right)\xi' + r \right\} \lambda' 
+2 +r^2 \phi'^2 + 2 \e^{2\lambda} \left(V r^2 -1 \right) \,, \\
\label{Eq2rr}
0=&\, 4\left\{ - 4 \left(1 -3\e^{-2\lambda} \right) \xi' +r \right\} \nu' + 2 - r^2 \phi'^2 -2 \e^{-2\lambda}+2 \e^{2\lambda} V r^2 \,, \\
\label{Eq2pp}
0 =&\, 2 \left(r + 8 \xi' \e^{-2\lambda} \right) \left( \nu'' + {\nu'}^2 \right) + 16 \xi'' \nu' \e^{-2\lambda} 
+ \left\{ -2 \left( r + 24 \xi' \e^{-2\lambda} \right) \lambda' + 2 \right\} \nu' 
-2 \lambda' + r\left( {\phi'}^2 + 2 \e^{2\lambda} V \right) \,. \\
\label{Eqphi2}
0=&\, - 8 \xi' \left(\e^{-2\lambda}-1 \right) \left( \nu'' + 2{\nu'}^2 \right) +\phi' \phi'' r^2 
 - 8 \nu' \xi' \left\{ \nu' \left(1-\e^{-2\lambda} \right) - \lambda' \left( 3\e^{-2\lambda} - 1 \right) \right\} \nonumber \\
&\, +r \left( \nu' r + 2 - \lambda' r \right)\phi'^2 - \e^{2\lambda} V' r^2 \,.
\end{align}
In Eqs.~(\ref{Eq2tt}), (\ref{Eq2rr}), (\ref{Eq2pp}), and (\ref{Eqphi2}), all the equations are not independent, for example, Eq.~(\ref{Eqphi2}) can be obtained 
from other equations. 
Therefore, in the following, we do not use Eq.~(\ref{Eqphi2}).

By combining Eq.~(\ref{Eq2tt}) with Eq.~(\ref{Eq2rr}), we obtain, 
\begin{align}
\label{V2}
V =&\, \frac{\e^{-2\lambda}}{r^2}\left\{ - 4 \left( \e^{-2\lambda}-1 \right) \xi'' - 4 \left( 1-3\e^{-2\lambda} \right) (\lambda' -\nu')
 \xi' +\e^{2\lambda} - 1 \right\} +\frac{\e^{-2\lambda}}{r} \left( \lambda' -\nu' \right) \, , \\
\label{xi2}
\phi' =&\, \pm \sqrt{ - \frac{8}{r^2} \left\{ \left( \e^{-2\lambda}-1 \right) \xi'' + \left( 1-3\e^{-2\lambda} \right) \left( \lambda' +\nu' \right) \xi'\right\} 
+\frac{2}{r} \left( \lambda' +\nu' \right) } \, . 
\end{align}
Furthermore, the combination of Eq.~(\ref{Eq2tt}) and Eq.~(\ref{Eq2pp}) gives,
\begin{align}
\label{f2}
0 =&\, - 8\, \left\{ \e^{-2\lambda} \left( \nu' r - 1 \right) +1 \right\} \xi'' - 8 \e^{-2\lambda} \left\{ r \left( \nu'' + {\nu'}^2 -3 \nu' \lambda' \right) 
+ \lambda' \left( 3 -\e^{2\lambda} \right) \right\} \xi' \nonumber \\
&\, -r^2 \left( \nu'' +{\nu'}^2 -\nu' \lambda' \right) -2r \left( \nu' +\lambda' \right) - \e^{2\lambda} +1 \, ,
\end{align}
which can be regarded with the differential equation for $\xi'$ and therefore for $\xi$ if $\nu=\nu(r)$ and $\lambda = \lambda(r)$ are given and the solution is 
\begin{align}
\label{f3}
\xi(r) = &\, - \frac{1}{8}\int \left[ \int \frac{
\e^{2\lambda} \left\{ \e^{2\lambda} + r^2 \left( \nu'' +{\nu'}^2 -\nu' \lambda' \right) +r (\nu' +\lambda') -1 \right\} }
{U \left(\nu' r - 1 + \e^{2\lambda} \right)} dr + c_1 \right] U dr +c_2
 \, , \nonumber \\
U(r) \equiv&\, \exp \left\{ -\int \frac{ r \left( \nu'' + {\nu'}^2 \right) + \lambda' \left( 3 - \e^{2\lambda} - 3 \nu' r \right)}
{\nu' r - 1 + \e^{2\lambda}} dr \right\} \, .
\end{align}
Here $c_1$ and $c_2$ are constants of the integration.

We may properly assume the profile of $\nu=\nu(r)$ and $\lambda = \lambda(r)$.
Therefore by using (\ref{f3}), we find the $r$-dependence of $\xi$, $\xi=\xi(r)$ and by using Eqs.~(\ref{V2}) and (\ref{xi2}),
we find the $r$ dependencies of $V$ and $\phi$, $V=V(r)$ and $\phi=\phi(r)$.
By solving $\phi=\phi(r)$ with respect to $r$, $r=r(\phi)$, we find $\xi$ and $V$ as functions of $\phi$,
$\xi(\phi)=\xi\left( r \left( \phi \right) \right)$, $V(\phi)=V\left( r \left( \phi \right) \right)$ which realise the model which has a solution
given by $\nu=\nu(r)$ and $\lambda = \lambda(r)$.

We should note, however, the expression of $\phi$ in (\ref{xi2}) gives a constraint, 
\begin{align}
\label{cons3}
 - \frac{8}{r^2} \left\{ \left( \e^{-2\lambda}-1 \right) \xi'' + \left( 1-3\e^{-2\lambda} \right) \left( \lambda' +\nu' \right) \xi'\right\} 
+\frac{2}{r} \left( \lambda' +\nu' \right) \geq 0 \, ,
\end{align}
so that the ghost could be avoided.
If Eq.~(\ref{cons3}) is not satisfied, the scalar field $\phi$ becomes pure imaginary.
We may define a new real scalar field $\zeta$ by $\phi=i\zeta$ $\left(i^2=-1\right)$ but because the coefficient
in front of the kinetic term of $\zeta$ becomes negative, $\zeta$ is a ghost, that is, a non-canonical scalar field.
The existence of the ghost generates the negative norm states in the quantum theory and therefore
the theory becomes inconsistent.


\subsection{Propagation of gravitational wave}

The conditions corresponding to (\ref{GWS2}) are now given by 
\begin{align}
\label{GWS2BH}
0=&\, \e^{-2\lambda} h_{rr} + \frac{1}{r^2}\sum_{i,j=1,2} {\tilde g}^{ij} h_{ij}\, , \quad 
0= \e^{-2\lambda} \left( \partial_r h_{rr} - 2\lambda' h_{rr} \right) + \sum_{i,j=1,2} \frac{{\tilde g}^{ij}}{r^2} \left( \partial_i h_{jr} 
+ \e^{-2\lambda} r {\tilde g}_{ij} h_{rr} - \frac{1}{r} h_{ji} \right) \, , \nonumber \\
0=&\, \e^{-2\lambda} \left( \partial_r h_{ri} - \lambda' h_{ri} - \frac{1}{r} h_{ri} \right) + \sum_{j,k=1,2} \frac{{\tilde g}^{jk}}{r^2} \left( \partial_j h_{ki} 
+ \e^{-2\lambda} r \left( {\tilde g}_{jk} h_{ri} + {\tilde g}_{ji} h_{kr} \right) - {\tilde\Gamma}^l_{jk} h_{li} - {\tilde\Gamma}^l_{ji} h_{kl} \right) \, .
\end{align}
We are also assuming $h_{t\mu}=h_{\mu t}=0$. 
Because we are considering the static and spherically symmetric spacetime, we may assume the scalar field $\phi$ and therefore $\xi=\xi(\phi)$ only depends on 
the radial coordinate $r$.
Then we find 
\begin{align}
\label{I2BH}
I^{(2)}_{\mu\nu} =&\, 2 \e^{-4\lambda} \left( \frac{d^2 \xi}{dr^2} - \lambda' \frac{d\xi}{dr} \right) \left\{ \nabla_\nu \nabla_r h_{r\mu} 
 - \nabla_\nu \nabla_\mu h_{rr} - {\nabla_r}^2 h_{\nu\mu} + \nabla_r \nabla_\mu h_{\nu r} \right\} \nonumber \\
&\, - 2 \e^{- 2\lambda -2\nu} \nu' \frac{d\xi}{dr} \left\{ \nabla_\nu \nabla_t h_{t\mu} 
 - \nabla_\nu \nabla_\mu h_{tt} - {\nabla_t}^2 h_{\nu\mu} + \nabla_t \nabla_\mu h_{\nu t} \right\} \nonumber \\
&\, + \frac{2 \e^{-2\lambda}}{r^3} \frac{d\xi}{dr} \sum_{i.j=1,2} {\tilde g}^{ij} \left\{ \nabla_\nu \nabla_i h_{j\mu} - \nabla_\nu \nabla_\mu h_{ji} 
 - \nabla_j \nabla_i h_{\nu\mu} + \nabla_j \nabla_\mu h_{\nu i} \right\} \, .
\end{align}
Now we consider the case that $h_{ir}=h_{ri}=0$. 
Then Eq.~(\ref{I2BH}) gives, 
\begin{align}
\label{I2BH2}
I^{(2)}_{ij} =&\, - 2 \e^{-4\lambda} \frac{d^2 \xi}{dr^2} {\nabla_r}^2 h_{ji} - 2 \e^{- 2\lambda -2\nu} \nu' \frac{d\xi}{dr} {\nabla_t}^2 h_{ji} 
+ \e^{-4\lambda}\lambda' \frac{d\xi}{dr} {\nabla_r}^2 h_{ji} - \frac{2 \e^{-2\lambda}}{r^3} \frac{d\xi}{dr} \sum_{k,l=1.2} {\tilde g}^{kl} \nabla_k \nabla_l h_{ij} \, .
\end{align}
In order $I^{(2)}_{ij}$ is proportional $\nabla^2 h_{ij}$, which is now given by 
\begin{align}
\label{I2BH3}
\nabla^2 h_{ij} =&\, - \e^{-2\nu}{\nabla_t}^2 h_{ij} + \e^{-2\lambda}{\nabla_r}^2 h_{ij} + \frac{1}{r^2} \sum_{k,l=1,2} {\tilde g}^{kl} \nabla_l \nabla_k h_{ij} \, ,
\end{align}
we need to require, 
\begin{align}
\label{I2BH4}
\nu' \frac{d\xi}{dr} = \frac{d^2 \xi}{dr^2} - 2 \lambda' \frac{d\xi}{dr} = \frac{1}{r} \frac{d\xi}{dr} \, ,
\end{align}
which gives strong constraints on the geometry if $\xi$ is not a constant. 
Especially, we find 
\begin{align}
\label{I2BH4B}
\nu = \ln \frac{r}{r_0}\, ,
\end{align}
where $r_0$ is a constant. 
Eq.~(\ref{I2BH4B}) tells that there is no horizon and therefore there is no solution for the black hole when the speed of the propagating speed 
exactly coincides with that of the light. 
This could tell that in the present universe, $\xi$ should be small even in the region where black holes exist. 

The conditions (\ref{I2BH4}), which are obtained by requiring that the propagation speed of gravitational wave could be identical 
with the speed of light in the static and spherically symmetric spacetime, are totally different from the condition (\ref{sgw1}) obtained in the FLRW spacetime 
although the condition $\nu' \frac{d\xi}{dr} = \frac{d^2 \xi}{dr^2} - 2 \lambda' \frac{d\xi}{dr}$ or $\frac{d^2 \xi}{dr^2} - 2 \lambda' \frac{d\xi}{dr} = \frac{1}{r} \frac{d\xi}{dr}$ 
in (\ref{I2BH4}) could correspond to the condition (\ref{sgw1}). 
The conditions are not compatible with each other. 
Especially, Eq.~(\ref{I2BH4B}) shows that the propagation speed of gravitational waves cannot be identical to the speed of light in the black hole background, 
that is, the conditions (\ref{I2BH4}) give constraints not only for the functional form of $\xi$ but for the spacetime geometry itself. 
Moreover, the change in the propagating speed generates lensing effects in addition to the usual gravitational lensing. 

The obstruction to construct the black hole solution in the single real scalar model could be similar to the present situation. 
In the case of the FLRW spacetime, we only need to specify the scale factor $a(t)$. 
Because the real scalar field has potential, there is an almost one-to-one correspondence between the scale factor and the potential \cite{Capozziello:2005tf}. 
In the case of the black hole, even if it is spherically symmetric, we need to specify two functions $g_{tt}$ and $g_{rr}$ (and more exactly we need to specify $g_{tr}$ to vanish) 
and one scalar theory cannot give a black hole solution with scalar hair but we need two scalar fields as in \cite{Nojiri:2020blr}. 
In the case of the sEGB gravity, the model is specified by two functions, that is, the scalar potential $V(\phi)$ and the coefficient function $\xi(\phi)$ 
of the Gauss-Bonnet term as in (\ref{g2}). 
This is the reason why we can construct the black hole solution in the Gauss-Bonnet gravity besides the problem of ghosts \cite{Nashed:2021cfs, Nojiri:2023qgd}. 
If we impose any constraint coming from the propagation speed of the gravitational wave, the number of independent functions reduces to one from two, 
which is not enough to realise two functions $g_{tt}$ and $g_{rr}$, which is a reason why we cannot construct the black hole solution under the condition (\ref{sgw1}). 
Hence, following the paper \cite{Nojiri:2020blr} it looks like we need two scalar fields to construct the black hole solution even in the case of the sEGB gravity.

\section{Summary and Discussion}

In this paper, general sEGB gravity which satisfies the condition that gravitational wave speed is equal to that of light is constructed. 
It is shown that such general theory may successfully describe the realistic inflation compatible with Planck experiment data.
It is also shown that the condition given by the requirement that the propagation speed of gravitational wave could be identical 
with the speed of light in the static and spherically symmetric spacetime, is totally different from the same condition obtained in the FLRW 
spacetime~\cite{Odintsov:2020xji, Odintsov:2020sqy, Oikonomou:2021kql, Oikonomou:2022ksx}. 
The corresponding conditions (\ref{I2BH4}) give constraints not only for the functional form of $\xi$ but for the spacetime geometry itself, i.e., equations that derive the black hole solution. 
Then sEGB black hole spacetime solution looks to be not compatible with the condition that gravitational wave speed is equal to light speed, i.e., with the GW170817 event. 
Nevertheless, the number of black holes between the earth and the point where the merger of the black hole and the neutron in the GW170817 event occurred, 
might be small and the cross-sections between the black hole and the gravitational wave could be also small. 
Then the majority of the gravitational waves reaching the earth could not be affected by the black holes and therefore the sEGB gravity could be 
consistent with GW17817 as long as the propagating speed of the gravitational wave is identical to that of light only in the FLRW spacetime background. 

If the propagating speed becomes smaller, the black hole works as a convex lens. 
On the other hand, if the speed becomes larger, the black hole works as a concave lens. 
Such effects might be observed in future gravitational wave experiments. 
Furthermore, if the speed of the gravitational wave is larger than that of light near the black hole horizon, the information inside the horizon may escape from the black hole. 
In any case, it is a challenge to find some scenario that can make the gravitational wave speed in sEGB black hole compatible with the speed of light, at least asymptotically

\section*{Acknoweledgements} 

This work was partially supported by MICINN (Spain), project PID2019-104397GB-I00 and by the program Unidad de Excelencia Maria de Maeztu CEX2020-001058-M, Spain (SDO). 
The work by SN was partially supported by the Maria de Maeztu Visiting Professorship at the Institute of Space Sciences, Barcelona.


\end{document}